\documentclass[
reprint,
10pt,
floatfix,
amsmath,
amssymb,
aps,
pra,
]{revtex4-2}

\usepackage{parskip}

\usepackage{graphicx}% Include figure files
\usepackage{dcolumn}% Align table columns on decimal point
\usepackage{bm}% bold math
\usepackage{color}
\usepackage{float}
\usepackage{caption}
\usepackage{ragged2e}
\begin{document}

\preprint{APS/123-QED}

\title{Heat-dissipation decomposition and free-energy generation\\ in a non-equilibrium dot with multi-electron states}% Force line breaks with \\
%\thanks{A footnote to the article title}%

\author{Chloe Salhani}
%\email{chloe.salhani@ntt.com}

\author{Kensaku Chida}%
\author{Takase Shimizu}%
\author{Toshiaki Hayashi}%
\affiliation{%
 Basic Research Laboratories, NTT Inc., Kanagawa, Japan
}%

\author{Katsuhiko Nishiguchi}%
 \email{nishiguchi-katsuhiko-kr@ynu.ac.jp}
 
\affiliation{%
 Institute for Multidisciplinary Sciences, Yokohama National University, Kanagawa, Japan
}%

\date{\today}% It is always \today, today,
             %  but any date may be explicitly specified

\begin{abstract}
We experimentally demonstrate  the decomposition of heat dissipation during free-energy generation in a nanometer-scale dot transitioning to a non-equilibrium steady state via single-electron counting statistics. An alternating-current signal driving a reservoir that injects multiple electrons into the dot makes it non-equilibrium, leading to free-energy generation, heat dissipation, and Shannon-entropy production. By analyzing the time-domain probability distributions of multi-electron states of the dot, we quantitatively decompose the heat dissipation into housekeeping and excess heats, thereby revealing their direct correlation with free-energy generation. This correlation suggests that the ratio of the generated free energy to the work applied to the dot, can potentially reach 0.5 under far-from-equilibrium conditions induced by a large signal, while an efficiency of 0.25 was experimentally achieved. These results establish a quantitative link between decomposed heat dissipation and free-energy generation in a multi-electron stochastic system, providing a thermodynamic framework for non-equilibrium electronic devices.  
\end{abstract}

%\keywords{Suggested keywords}%Use showkeys class option if keyword
                              %display desired
\maketitle

%\tableofcontents

\section{\label{sec:level1}introduction}

Thermodynamics provides essential insights into the energetic limits of physical systems, such as the Landauer bound on information processing \cite{Landauer1961}. 
Experimental tests in microscopic platforms have confirmed these principles primarily under near-equilibrium conditions \cite{Berut2012,Jun2014,Koski2014_szilard,Berut2015,Martini2016,Hong2016,Barker2022}. In contrast, electronic devices typically operate far from equilibrium to achieve high speed, output, and efficiency \cite{Shiraishi2016,Wolpert2019,Proesmans2020,Lee2022}, motivating a corresponding extension of energetic limits to this regime.  

A non-equilibrium state arises when an external force breaks global detailed balance and produces entropy, leading to free-energy generation as a key thermodynamic output. Within stochastic thermodynamics, this entropy production consists of heat dissipation and a change in Shannon entropy, which encodes the information content of the system and contributes to non-equilibrium free energy \cite{Esposito2011}. 
To analyze transitions between non-equilibrium steady states (NESSs), it is further useful to decompose the entropy production into excess and housekeeping parts: the former reflects the energetic cost to drive the system away from a steady state, while the latter quantifies the dissipation required to sustain a NESS \cite{Oono1998,Hatano2001,Esposito2007,Yoshimura2023}. 
This decomposition underpins non-equilibrium inequalities and fluctuation relations \cite{Hatano2001,Esposito2007,Seifert2005,Gaspard2004,Speck2005,Broeck2010,Esposito2010}. However, how the decomposed contributions quantitatively relate to free-energy generation remains experimentally unexplored. 

Electronic devices provide well-controlled experiments on non-equilibrium physics: in particular, single-electron devices composed of quantum dots have enabled precise measurements of work and heat under driven conditions \cite{Koski2014_szilard,Barker2022,Averin2011,Kung2012,Saira2012,Pekola2012,Hofmann2016,Hofmann2017}. In these systems, free-energy differences are typically inferred via fluctuation relations such as the Jarzynski equality, yielding equilibrium state functions despite non-equilibrium driving. 
Moreover, the accessible charge states are restricted to $N=0$ or $1$ by large charging energies $E_{\rm C}\gg k_{\rm B}T$, limiting the Shannon entropy to $\ln 2$ and  the accessible range of free-energy variation. Here, $k_{\rm B}$ is the Boltzmann constant, and $T$ is the temperature. In addition, since these systems are typically not operated under steady non-equilibrium driving, an experimental decomposition of heat into excess and housekeeping components has not yet been achieved, and would, in any case, remain limited by the two-state nature. Related studies in non-electronic platforms such as colloidal beads and granular media have verified steady-state relations \cite{Trepagnier2004,Mounier2012,Ando2024}, but they also do not provide explicit heat decomposition or probe the multi-electron dynamics. 

In contrast, multi-electron systems operating at $E_{\rm C}\lesssim k_{\rm B}T$ possess a large state space and allow strong non-equilibrium driving.
While equilibrium electron-number distributions are Gaussian \cite{Nishiguchi2014Nanotech,Carles2015}, non-equilibrium driving can induce non-Gaussian distortions, leading to qualitatively different entropy production and free-energy characteristics.
This expanded state space enables simultaneous and pronounced variations of excess and housekeeping heat, necessitating an analysis based on the full time-dependent probability distribution.

In this study, we experimentally quantify heat dissipation and Shannon entropy in a small dot with multi-electron occupancy driven by an AC signal. Using single-electron counting statistics, we track the dynamics of many electrons in a device that mimics a dynamic random-access-memory (DRAM) cell. Although $E_{\rm C}\lesssim k_{\rm B}T$, the electron number remains a well-defined state variable.
We decompose the total heat into housekeeping and excess components, reveal their quantitative relation to non-equilibrium free-energy generation and efficiency, and demonstrate how multi-electron systems provide a direct thermodynamic framework connecting heat dissipation to electronic-device operation far from equilibrium.

\section{\label{sec:experimental-system}Experimental system}

\begin{figure}[b]
\includegraphics[width=\columnwidth]{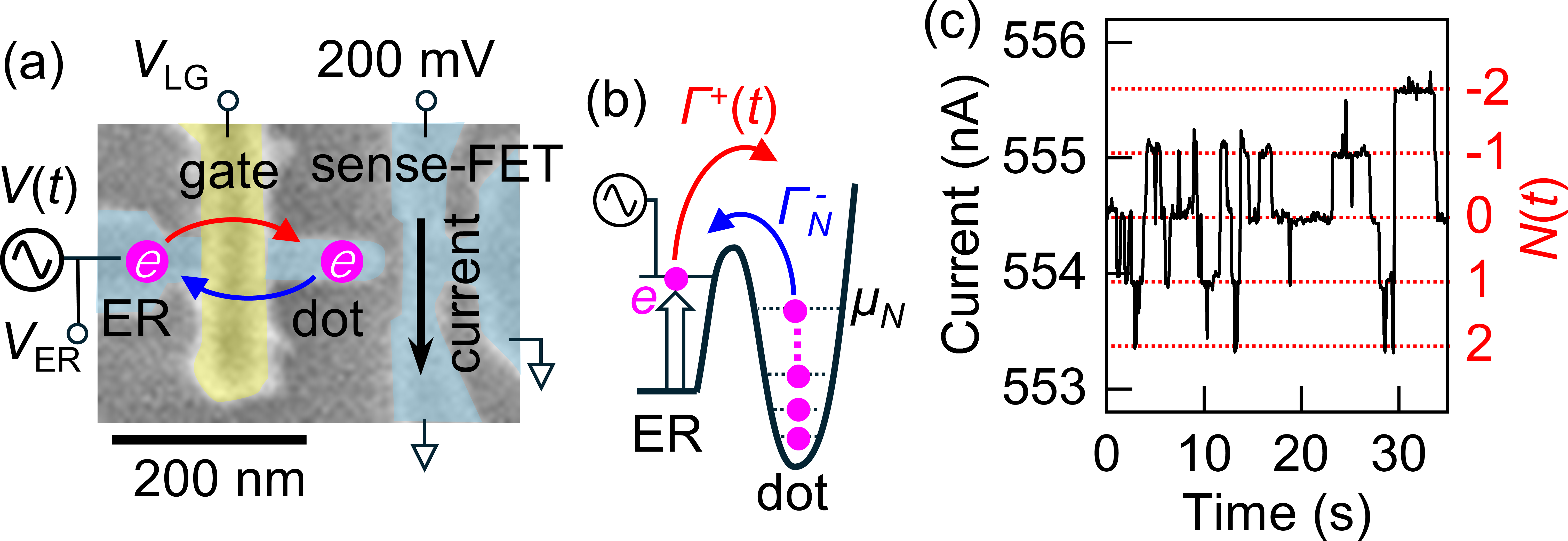}
\caption{\justifying \label{Figure 1} (a) False scanning-electron-microscope image of the device for single-electron counting statistics. The constant ER voltage $V_{\rm ER}$ was 0.5 V. The details of the device structure are explained in Supplemental Material.
(b) Energy band diagram between the ER and dot when the excess number of electrons in the dot is $N$. The Fermi energy of the ER is modulated by an AC signal $V(t)$ superimposed on $V_{\rm ER}$. (c) Current flowing through the sense-FET without the AC signal. The sampling rate of the sense-FET was 20 Hz.
}\end{figure}

Figure \ref{Figure 1}(a) shows our device for single-electron counting statistics: the individual motion of single electrons traveling back and forth between an electron reservoir (ER)  and the dot is monitored.\cite{Nishiguchi2008JJAP} Since all measurements were carried out at room temperature, the electron motion originates not from tunneling events but from thermal hopping over an energy barrier under a gate [see Fig. 1(b)].\cite{Nishiguchi2011APL} These structures and principles are similar to conventional DRAMs, except that the dot in our device is very small (about 10 attofarads in capacitance), storing approximately twenty electrons. The electrons in the dot are counted with a field-effect transistor (FET), referred to as the sense-FET: the current flowing through the sense-FET increases/decreases in a step-like pattern when a single electron leaves/enters the dot, as shown in Fig. \ref{Figure 1}(c). From the step height caused by a single electron and the sense-FET's transconductance, the charging energy $E_{\rm C}$ for one electron stored in the dot was estimated to be 9.8 meV. \cite{Nishiguchi2008JJAP}

We control equilibrium and non-equilibrium states between the ER and the dot, hereafter defined as the ER-dot system, by modulating the Fermi energy of the ER.\cite{Salhani2024} In the equilibrium state, constant voltages are applied to the ER and gate, respectively. Since the energy barrier height controlled by the gate governs the transition rate of electrons surmounting the barrier considering Kramers' rate,\cite{Kramers1940} the transition rate $\varGamma_0$ for an electron to enter the dot is constant, while the rate $\varGamma_N^{-}$ for an electron to leave the dot is given by $\varGamma_0 \exp(\beta \mu_N)$,\cite{Nishiguchi2014Nanotech} where $\mu_N=2E_{\rm C}(N-0.5)$ is the chemical potential of the dot, $N$ is the deviation from the average number of electrons in the dot at equilibrium [see Fig. \ref{Figure 1}(b)], and $\beta=(k_{\rm B}T)^{-1}$ is the inverse temperature. These transition rates are estimated experimentally from $N(t)$, where $N(t)$ is the time evolution of  $N$ [Fig. \ref{Figure 1}(c)].\cite{Nishiguchi2014Nanotech} $\varGamma_0$ is adjusted to be 0.1 Hz to ensure the electron motion is slow enough to be precisely monitored.

The non-equilibrium state is induced by superimposing an AC voltage signal,  $V(t)=S_{\rm AC}\sin(\omega_{\rm AC}t+\phi)$, on the constant ER voltage.\cite{Salhani2024} Here, $S_{\rm AC}$, $\omega_{\rm AC}/2\pi$, and $\phi$ are the amplitude, frequency, and phase of the AC signal. 
The AC signal changes the energy barrier for electrons in the ER by $-eV(t)$, and therefore the transition rate for entering the dot is modulated over time,  $\varGamma^+(t)=\varGamma_0 \exp[-\beta eV(t)]$. In contrast,  the escape rate $\varGamma_N^{-}=\varGamma_0 \exp(\beta \mu_N)$ remains unaffected. 
In this study, $\omega_{\rm AC}/2\pi$ of 1 Hz larger than $\varGamma_0 $ prevents electrons from following the AC signal and thus pushes the ER-dot system out of equilibrium. Detailed frequency dependence is discussed in \cite{Salhani2024} and Appendix \ref{ap:MEsolution}.

To perform statistical analysis, all measured quantities are evaluated as phase-averaged ensemble averages. In the experiment, the phase $\phi$ of the applied AC signal is randomized for each measurement, and the results are averaged over many repetitions. This procedure removes phase-dependent contributions and provides access to the intrinsic stochastic properties of the system.

A central quantity in our analysis is the time-dependent probability $\rho_N(t)$, which we obtain from the statistical analysis of repeated measurements. Here, $\rho_N(t)$ represents the probability that the dot contains an excess electron number $N$ at time $t$. The probability distribution is obtained by converting the measured sense-FET current into$N(t)$ and averaging over the phase ensemble.

In the theoretical analysis, we assume that the phase $\phi$ is uniformly distributed. Under this assumption, all observables correspond to phase-averaged quantities, and the time-averaged driving voltage vanishes. This ensures direct consistency between experimental observables and the theoretical framework.

\section{Theoretical framework}
\subsection{\label{sec:ME}Master equation}

The dynamics of the ER-dot system can be described by the following master equation:
\begin{align}
    &\dot{\rho}_N(t)=\varGamma^{+}(t) \rho_{N-1}(t)\notag \\
    &\hspace{16mm} -[\varGamma^{+}(t)+\varGamma_N^{-}]\rho_N(t)+\varGamma_{N+1}^{-}\rho_{N+1}(t), \label{eq:ME}
\end{align}
where $\sum_{N=-\infty}^\infty \rho_N(t) =1$. Throughout this paper, the dot notation $\dot{\Box}$ represents the time derivative of $\Box$.

The Fourier series expansion of the time-dependent quantities $\varGamma^{+}(t)$ and $\rho_N(t)$ provides a systematic framework to analyze the master equation and extract its dominant dynamics (see Appendix \ref{ap:MEsolution}). Based on this analysis, the time-dependent transition rate $\varGamma^{+}(t)$ can be approximated by its time-averaged value,
\begin{equation}
    \varGamma^{+}(t)\approx \overline{\varGamma^+}=\varGamma_0 I_0(\beta eS_{\rm AC}), \label{eq:bar_Gamma}
\end{equation}
where $I_0(\Box)$ is the modified Bessel function of the first kind.

Under this approximation, which effectively corresponds to a coarse-grained description over the driving period, the probability distribution $\rho_N(t)$ in the NESS can be approximated by a stationary distribution $\rho_N^{\rm SS}$, which takes a Gaussian form, as obtained from the Fourier analysis (see Appendix \ref{ap:MEsolution}):
\begin{align}
     \rho_N^{\rm SS}=Z^{-1} \exp \left[ -\beta E_{\rm C}(N-\langle N_{\rm SS}\rangle)^2 \right], \label{eq:rho2}
\end{align}
where 
\begin{align}
    \langle N_{\rm SS}\rangle=\ln (\overline{\varGamma^+}/\varGamma_0)/(2\beta E_{\rm C}) \label{eq:N_ss}
\end{align}
corresponds to the zero-frequency component in the Fourier expansion of $\langle N(t)\rangle$ in the NESS. $Z$ is given by $\sum_{N=-\infty}^{\infty} {\rm exp} \left[ -\beta E_{\rm C}(N-\langle N_{\rm SS}\rangle)^2 \right]$.

The validity of these approximations will be examined by direct comparison with the experimental results presented in Sec.~\ref{sec:p-dynamics}. 
This approach is further supported by previous studies on stochastic pumps,\cite{Raz2016} where similar approximations have been successfully applied to AC-driven systems with modulation faster than the experimental sampling rate, making it particularly suitable for the present system.

\subsection{Heat dissipation rate for single-electron motion}

Heat dissipation is estimated from individual electron motions between the ER and dot. When the excess electron number in the dot changes from $N$ to $N+1$ or $N-1$, its heat dissipation $Q_N^{+1}$ or $Q_N^{-1}$, respectively, is given by 
\begin{equation}
    Q_N^{\pm 1}=-(H_{N \pm 1}-H_N). \label{eq:Qpm1}
\end{equation}
Here, $H_N$ is the Hamiltonian when the number of excess electrons in the dot is $N$:\cite{Pekola2012} $H_N=E_{\rm C}(N-N_{\rm off})^2-E_{\rm C}N_{\rm off}^2$, where $N_{\rm off}=-eV(t)/(2E_{\rm C})$. Consequently, Eq. (\ref{eq:Qpm1}) yields [see Fig. \ref{Figure 1}(b)]
\begin{align}
  &Q_N^{+1}=-eV(t)-\mu_{N+1},  {\rm and}\label{eq:Qp1}\\
  &Q_N^{-1}=eV(t)+\mu_{N}. \label{eq:Qm1}
\end{align}

By multiplying Eqs. ({\ref{eq:Qp1}}) and  ({\ref{eq:Qm1}}) by probability flows, the heat-dissipation rates for electrons entering and leaving the dot are given by 
\begin{align}
    &\langle \dot{Q}^{+}(t)\rangle=\sum_{N=-\infty}^{\infty} \rho_N(t)\varGamma^{+}(t) [-eV(t)-\mu_{N+1}]  \text{ and} \label{eq:Qin_1}\\
   &\langle \dot{Q}^{-}(t)\rangle=\sum_{N=-\infty}^{\infty} \rho_N(t)\varGamma_N^{-}[eV(t)+\mu_N] \label{eq:Qout_1}, 
\end{align}
respectively. 
The total heat dissipation rate $\langle\dot{Q}_{\rm T}(t)\rangle$ is defined as $\langle\dot{Q}_{\rm T}(t)\rangle=\langle\dot{Q}^{+}(t)\rangle+\langle\dot{Q}^{-}(t)\rangle$. 

Next, we consider heat dissipation in a NESS. $\rho_N(t)$ in the NESS relaxes to a Gaussian distribution $\rho_N^{\rm SS}$ given by Eq. (\ref{eq:rho2}). Given that $\mu_N$, $\rho_N^{\rm SS}$, and $\varGamma_N^{-}$ are time-independent, that $\varGamma^{+}(t)$ is averaged over time, i.e., $\overline{\varGamma^+}$, and that the phase or time-averaged $V(t)$ is zero, Eqs.  (\ref{eq:Qin_1}) and (\ref{eq:Qout_1}) yield  $\langle\dot{Q}^{+}(t)\rangle$  and  $\langle\dot{Q}^{-}(t)\rangle$  in a NESS:
\begin{align}
&\langle \dot{Q}^{+ \rm SS}(t)\rangle=-eV(t)\varGamma^{+}(t)-2E_{\rm C}(\langle N_{\rm SS}\rangle+0.5)\overline{\varGamma^+}, \label{eq:Qin_3}\\
&\langle\dot{Q}^{-\rm SS}(t)\rangle=2E_{\rm C}(\langle N_{\rm SS}\rangle+0.5)\overline{\varGamma^+}\label{eq:Qout_3}
\end{align}
Because the instantaneous dissipation depends on the phase of the AC drive, the term $eV(t)\varGamma^+(t)$ is left as a time-dependent quantity here, while its time-averaged form is given later (Appendix \ref{ap:HK}). Consequently, $\langle\dot{Q}_{\rm T}(t)\rangle$ in a NESS becomes
\begin{equation}
   \langle\dot{Q}_{\rm T}^{\rm SS}(t)\rangle=\langle\dot{Q}^{+ \rm SS}(t)\rangle+\langle\dot{Q}^{- \rm SS}(t)\rangle =-eV(t)\varGamma^{+}(t). 
   \label{eq:Qt_SS}
\end{equation}
Since $-e\varGamma^{+}(t)$ corresponds to electric current from the ER to the dot, Eq. (\ref{eq:Qt_SS}) corresponds to the electric power dissipation or work rate to maintain the ER-dot system in the NESS.

\subsection{\label{sec:heat-decomposition-theory}Decomposition into excess and housekeeping heat}

Excess heats $Q_{{\rm EX,}N}^{+1}$ and $Q_{{\rm EX,}N}^{-1}$ for an electron entering and leaving the dot when the excess electron number is $N$ are given by\cite{Hatano2001}
\begin{equation}
    \beta Q_{{\rm EX,}N}^{\pm 1}=\ln \left(\frac{\rho_{N \pm 1}^{\rm SS}}{\rho_N^{\rm SS}} \right). \label{eq:Qex} 
\end{equation}
Substituting Eq. (\ref{eq:rho2}) into equation Eq. (\ref{eq:Qex}) gives
\begin{align}
     &Q_{{\rm EX,}N}^{+1}=2E_{\rm C}\langle N_{\rm SS}\rangle-\mu_{N+1}, \text{ and} \label{eq:Qex_p1}\\
     &Q_{{\rm EX,}N}^{-1}=-2E_{\rm C}\langle N_{\rm SS}\rangle+\mu_{N}. \label{eq:Qex_m1}
\end{align}
Housekeeping heats $Q_{{\rm HK,}N}^{+1}$ and $Q_{{\rm HK,}N}^{-1}$ for an electron entering and leaving the dot, respectively, at $N$ follow 
\begin{equation}
\begin{split}
       \beta Q_{{\rm HK,}N}^{\pm 1}&=\ln \left[\frac{R(N \pm 1|N)}{R(N|N \pm 1)}\frac{\rho_N^{\rm SS}}{\rho_{N \pm 1}^{\rm SS}} \right] \\
       &=\pm [-eV(t)-2E_{\rm C}\langle N_{\rm SS}\rangle], \label{eq:Qhk}  
\end{split}
\end{equation}
where $R(y|x)$ is the transition rate for $N$ to change from $x$ to $y$. Consequently, Eqs. (\ref{eq:Qp1}) and (\ref{eq:Qm1}) are decomposed:
\begin{align}
  &Q_N^{+1}=[-eV(t)-2E_{\rm C}\langle N_{\rm SS}\rangle]\notag\\
  &\hspace{20mm}+[2E_{\rm C}\langle N_{\rm SS}\rangle-\mu_{N+1}] \text{ and} \label{eq:Qp1d}\\
  &Q_N^{-1}=[eV(t)+2E_{\rm C}\langle N_{\rm SS}\rangle]\notag\\
  &\hspace{20mm}+[-2E_{\rm C}\langle N_{\rm SS}\rangle+\mu_{N}]. \label{eq:Qm1d}
\end{align}
The first and second square brackets on the right-hand sides correspond to housekeeping and excess heats, respectively.

Equations~(\ref{eq:Qp1d}) and (\ref{eq:Qm1d}) clarify the connection between the decomposition at the level of single-electron transitions and the averaged heat dissipation rates in Eqs.~(\ref{eq:Qin_3}), (\ref{eq:Qout_3}), and (\ref{eq:Qt_SS}). 
Within this correspondence, the physical origin of the factor of $1/2$ in Eqs.~(\ref{eq:Qin_3}) and (\ref{eq:Qout_3}) can be traced back to the offset in the electrochemical potential $\mu_{N+1}=2E_{\rm C}(N+0.5)$, which reflects the discrete nature of charge transitions. 
Upon ensemble averaging, this offset gives rise to a constant contribution that contributes to the excess heat in the thermodynamic decomposition.
These expressions also provide a direct route to experimentally reconstruct $Q_{\rm EX}$ and $Q_{\rm HK}$ from the time evolution of $N$, as discussed below.

\section{\label{sec:p-dynamics}Dynamics of the probability distribution and entropy production}

\begin{figure}
\includegraphics[width=\columnwidth]
{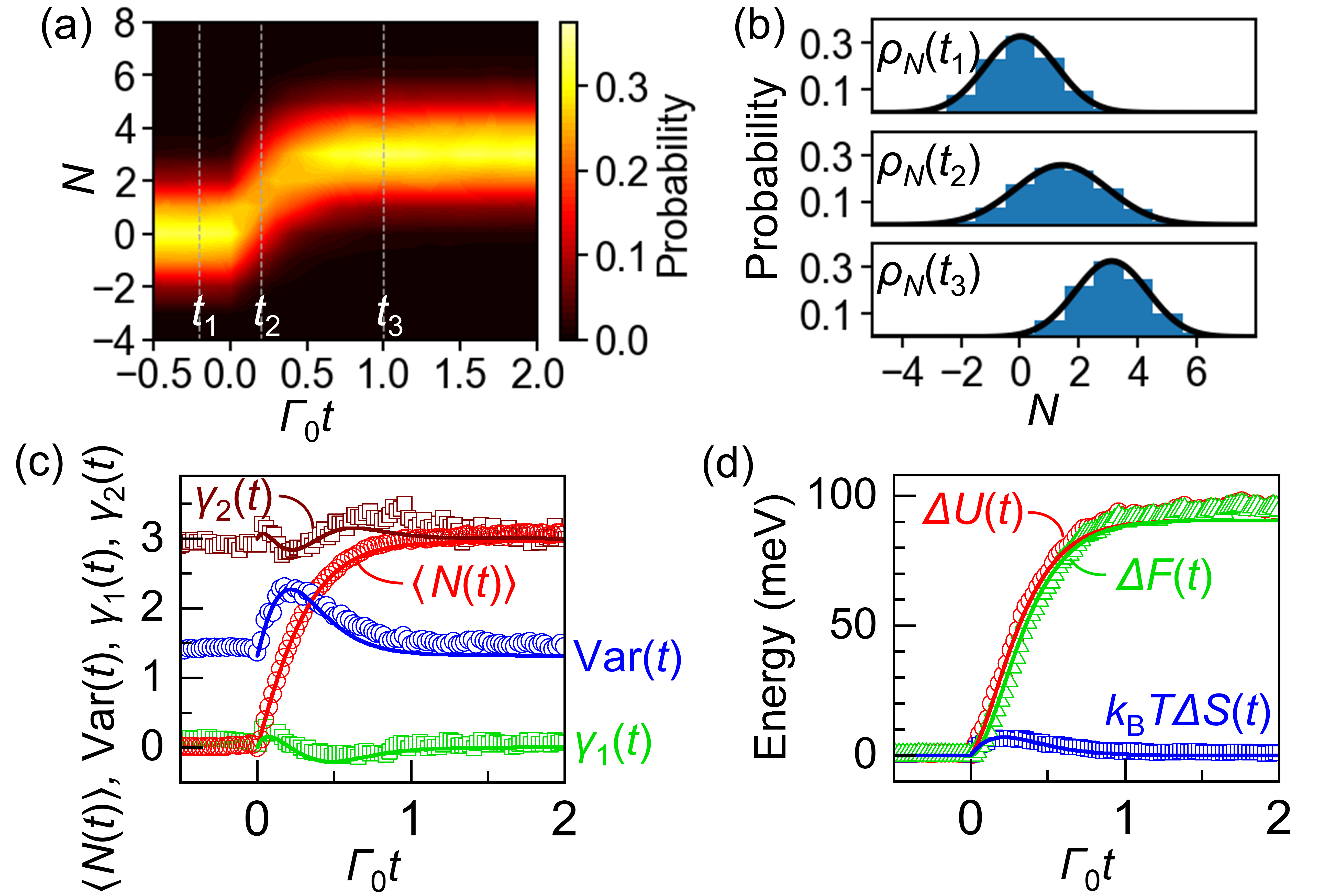}
\caption{\justifying \label{Figure 2} (a) Contour plot of the $\rho_N (t)$ distributions from an equilibrium state to an  NESS. The AC signal with amplitude $S_{\rm AC}$ of 100 mV was applied at $t=0$. Experimental data were averaged over 3000 repetitions. (b) $\rho_N (t)$ distributions at $t_1$, $t_2$, and $t_3$ depicted in (a). The solid lines are the Gaussian fits to the experimental results. (c) Transient characteristics of the average $\langle N(t) \rangle$, variance ${\rm Var}(t)$, skewness $\gamma_1(t)$, and kurtosis $\gamma_2(t)$ of $\rho_N (t)$. (d)  Transient characteristics of the deviations of the internal energy $\Delta U(t)$, free energy $\Delta F(t)$, and Shannon entropy $k_{\rm B}T\Delta S(t)$ from their initial values at $t=0$. Open marks are the experimental data, and the solid lines are the numerical results of the master equation.}
\end{figure}

Figures \ref{Figure 2}(a) and (b) show the distributions of $\rho_N (t)$ over time $t$ when an AC signal was applied at $t=0$. To characterize equilibrium and non-equilibrium states of the ER-dot system, we consider the average  $\langle N(t) \rangle$, given by $\sum_{N=-\infty}^\infty \rho_N(t)N$, and the variance ${\rm Var}(t)$, given by $\sum_{N=-\infty}^\infty \rho_N(t)[N-\langle N(t) \rangle]^2$, of $N$ at time $t$, which are summarized in Fig. \ref{Figure 2}(c).

Before the AC signal is applied, the system is in equilibrium: $\rho_N(t)$ follows a Gaussian distribution with a mean  $\langle N(t) \rangle$ of zero and a variance ${\rm Var}(t)$ given by $(2\beta E_{\rm C})^{-1}$ as shown by the $\rho_N(t_1)$ distribution in Fig. 2(b), meaning the electron motion between the ER and the dot is thermally-activated.\cite{Nishiguchi2014Nanotech}  
After the AC signal is applied, the dynamic change in the ER voltage causes a temporal disruption of the detailed balance condition, placing the ER-dot system in non-equilibrium, which generates entropy production as discussed below. 
As shown in Fig. {\ref{Figure 2}}(b), the bell-shaped distribution of $\rho_N (t)$ shifts positively in $N$ and $\langle N(t) \rangle$ increases over time. Meanwhile, ${\rm Var}(t)$ initially increases and then decreases. During this interval, the skewness $\gamma_1(t)$ and kurtosis $\gamma_2(t)$ of the $\rho_N (t_2)$ distribution deviate from 0 and 3, respectively, meaning that $\rho_N (t_2)$ deviates from a Gaussian distribution. 
Finally, after $t \sim 1/\varGamma_0$, the distribution $\rho_N(t_3)$ stabilizes to a Gaussian distribution: its average $\langle N(t) \rangle$ and variance ${\rm Var}(t)$ are given by $\langle N_{\rm SS} \rangle$ and $(2\beta E_{\rm C})^{-1}$, respectively. The transition rate for electrons entering the dot balances with the rate of escaping electrons, meaning the local detailed balance holds between the ER and the dot, and the system reaches a NESS.\cite{Salhani2024} These features mean that $\rho_N(t_3)$ can be represented by the stationary probability distribution $\rho_N^{\rm SS}$ given by Eq. (\ref{eq:rho2}) derived from the Fourier analysis of the master equation (also see Appendix \ref{ap:MEsolution}). 

As shown in  Fig. {\ref{Figure 2}}(c), the dynamic features of $\rho_N(t)$ are traced well by the numerical results of the master equation using the Fourier series expansion, implying that the  approximation by such Fourier analysis described in Sec. \ref{sec:ME} is valid for the ER-dot system driven by an AC signal with a frequency higher than $\varGamma_0$ [Figs. 2-4].

Next, we discuss the entropy production during the dynamic ER-dot system. The entropy production in non-equilibrium can be expressed as the sum of the change in the Shannon entropy $S(t)$ and the heat dissipation. First, we discuss $S(t)$ and its change $\Delta S(t)$ given by $-\sum_{N=-\infty}^\infty \rho_N(t)\ln \rho_N(t)$ and $S(t)-S(0)$, respectively. As shown in Figs. \ref{Figure 2}(c) and (d), during the state transition from the equilibrium state to the NESS, ${\rm Var}(t)$  increases, leading to an increase in $\Delta S(t)$. In the NESS, ${\rm Var}(t)$ returns to its  initial value $(2\beta E_{\rm C})^{-1}$, as confirmed at various values of $\omega_{\rm AC}$ and $S_{\rm AC}$ in Ref. \cite{Salhani2024}, implying $\Delta S(t)=0$. Since $k_{\rm B}T\Delta S(t)$ is much smaller than the heat dissipation explained in the next section, we  consider the entropy production to be dominated by heat dissipation in the present system. 

Additionally, $\Delta S(t)$ is sufficiently small that the change in the internal energy $\Delta U(t)$, given by $\sum_N E_{\rm C} N^2 \rho_N(t)$, is close to the free-energy change $\Delta F(t)$ [see Fig.~\ref{Figure 2}(c)], since $\Delta F(t)=\Delta U(t)-k_{\rm B}T\Delta S(t)$.\cite{Esposito2011}
This behavior can be understood from the near-Gaussian form of $\rho_N(t)$: the skewness $\gamma_1(t)$ and kurtosis $\gamma_2(t)$ remain close to 0 and 3, respectively [Fig.~\ref{Figure 2}(c)], indicating only a small deviation from a Gaussian distribution. Under this condition, $\Delta F(t)$ can be approximated as $E_{\rm C}\langle N(t)\rangle^2$ (see Appendix~\ref{ap:MEsolution}). Consistent with this approximation, as shown in Fig.~\ref{Figure 2}(d), $\Delta F(t)$ increases monotonically and saturates at $E_{\rm C}\langle N_{\rm SS}\rangle^2$ in the NESS.

\section{\label{sec:heat-decomposition-experiment} Experimental decomposition of heat dissipation}

We next evaluate the heat dissipation in the dynamic ER-dot system and examine its decomposition based on the theoretical framework introduced in Sec.~\ref{sec:heat-decomposition-theory}.

\begin{figure}[b]
\includegraphics{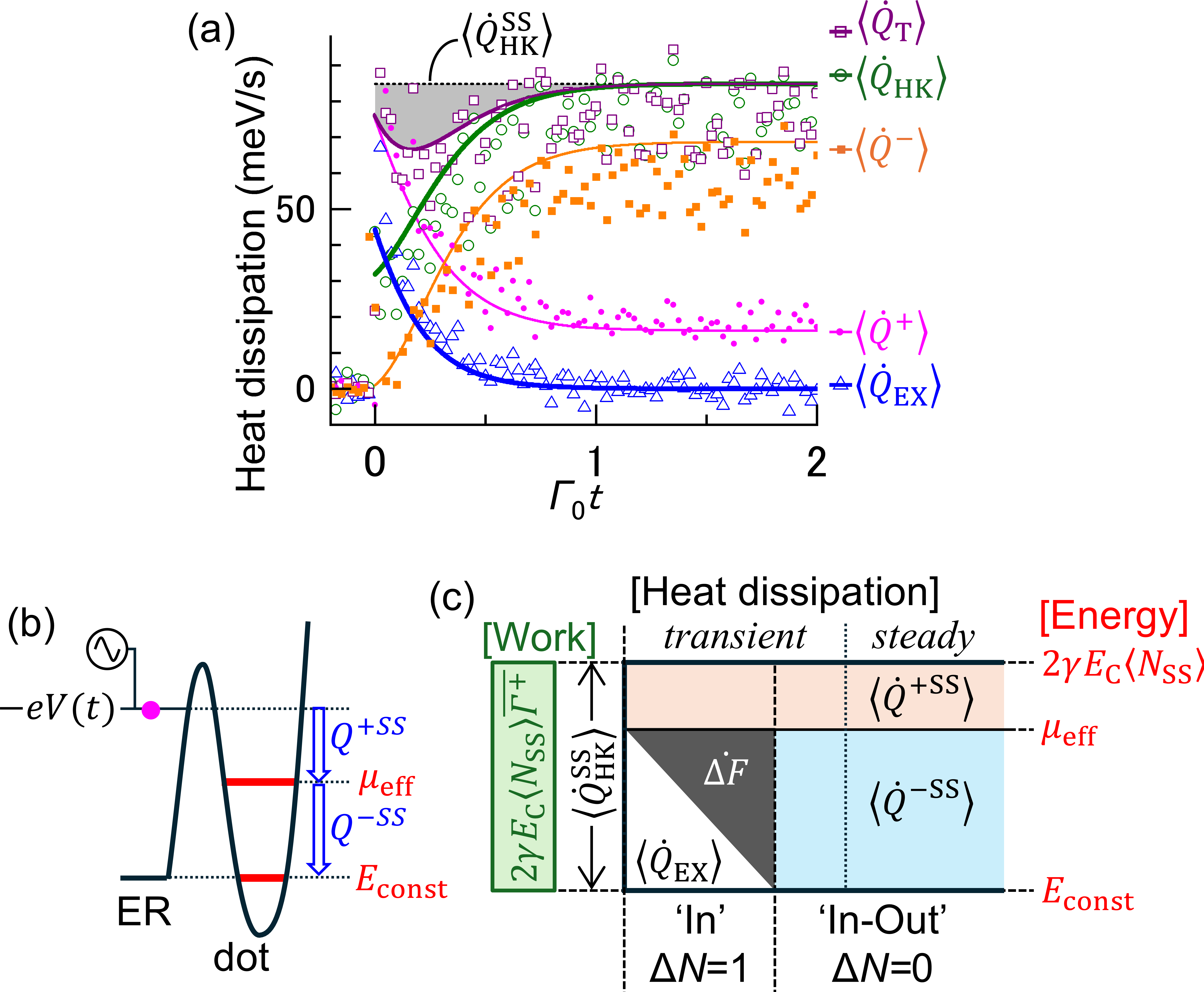}
\caption{\justifying \label{Figure 3} (a) $\langle\dot{Q}_{\rm T}\rangle$ and its decomposition into $\langle\dot{Q}_{\rm EX}\rangle$, $\langle\dot{Q}_{\rm HK}\rangle$, $\langle\dot{Q}^{+}\rangle$, and $\langle\dot{Q}^{-}\rangle$. For brevity, the explicit time argument $(t)$ of these quantities is omitted. $Q_{\rm EX}$ and $Q_{\rm HK}$ are evaluated from $N(t)$ [Fig. 1(c)] using Eqs. (14)–(16) together with the experimentally determined transition rates. The steady-state quantities $\rho_N^{\rm SS}$ and $\langle N_{\rm SS} \rangle$ are obtained from the experimental data in the NESS (Fig. 2). The corresponding heat dissipation rates are then derived by taking the time derivative. Scatter points are experimental averages over 3000 repetitions, and solid lines are numerical results from the master equation. The gray region shows $\langle\dot{Q}_{\rm HK}^{\rm SS}\rangle - \langle\dot{Q}_{\rm T}\rangle$. $S_{\rm AC}=100~{\rm mV}$. 
(b) Energy band diagram of the ER–dot system.  $Q^{\rm +SS}$ and $Q^{\rm -SS}$ correspond to heat dissipation when one electron enters and leaves, respectively, the dot in the NESS [see Eqs. (\ref{eq:Qin_1}) and (\ref{eq:Qout_1})].
(c) Schematic representation of heat dissipation, free-energy generation, and work. The left and right vertical axes represent these rates and the corresponding energy levels (in red), respectively. The horizontal axis indicates “In’’ and “In–Out’’ events, in which $N$ increases by one and in which the same number of electrons enter and leave the dot, respectively. Although these events occur stochastically, they are grouped into two categories. The shaded areas represent the total heat dissipated in each process.}
\end{figure}

We first present the experimental results.  Figure \ref{Figure 3}(a) shows the time evolution of $\langle\dot{Q}_{\rm T}(t)\rangle$ and its decomposition into $\langle\dot{Q}_{\rm EX}(t)\rangle$ and $\langle\dot{Q}_{\rm HK}(t)\rangle$. $Q_{\rm EX}$ and $Q_{\rm HK}$ are experimentally reconstructed from $N(t)$ using Eqs. (14)–(16) together with the measured transition rates. At $t=0$,  $\langle\dot{Q}_{\rm EX}(t)\rangle$ is maximal. Then it monotonically decreases and becomes zero in the NESS.  In contrast, $\langle\dot{Q}_{\rm HK}(t)\rangle$ is already finite in the transient regime and gradually  approaches a constant value $\langle\dot{Q}_{\rm HK}^{\rm SS}\rangle$.

The experimental analysis on $\langle\dot{Q}_{\rm T}(t)\rangle$ and $\langle\dot{Q}_{\rm HK}^{\rm SS}\rangle$, combined with the numerical analysis using the master equation of Eq. (\ref{eq:ME}), leads to the following energy balance relation:
\begin{equation}
    \int_0^{\tau_{\rm SS}} \langle\dot{Q}_{\rm T}(t)\rangle dt + \Delta F_{\rm SS} = \int_0^{\tau_{\rm SS}} \langle\dot{Q}_{\rm HK}^{\rm SS}\rangle dt, \label{eq:Qt+dF=Qhk}
\end{equation}
which connects the transient heat dissipation to the free energy $\Delta F_{\rm SS}$ in the NESS. 

To understand the physical origin of this relation, we now consider the microscopic energy or heat flows. Figure 3(b) provides a schematic overview of the heat dissipation processes. In the NESS, Eqs. (\ref{eq:Qin_3}) and (\ref{eq:Qout_3}) describe the heat flow required to maintain the steady state. The term $2E_{\rm C}\langle N_{\rm SS} \rangle$ can be interpreted as an effective chemical potential $\mu_{\rm eff}$, which characterizes the energy level of the dot. Equation~(\ref{eq:Qin_3}) represents the heat dissipation associated with an electron entering the dot from the ER with energy $-eV(t)$  [see $Q^{\rm +SS}$ shown in the left panel of Fig. ~\ref{Figure 3}(b)], while Eq.~(\ref{eq:Qout_3}) describes the dissipation when an electron in the dot returns to the ER with the reference energy $E_{\rm const}$ (see $Q^{\rm -SS}$). $E_{\rm const}$ corresponds to the Fermi energy of the ER in the absence of the AC signal. These kinds of heat dissipation stem from  the work applied to the system, i.e., the time-dependent potential $-eV(t)$ . 

These considerations on heat flows are extended to those in the transition state before the NESS by classifying stochastic events into two types: “In” events and “In–Out” events [see Fig. \ref{Figure 3}(c)].   “In” events in which an electron only enters the dot contribute the increase in $\langle N(t) \rangle$ from zero to $\langle N_{\rm SS}\rangle$. An electron entering the dot dissipates heat given by $Q_N^{+1}=[-eV(t)-\mu_{\rm eff}] +(\mu_{\rm eff} -\mu_{N+1})$ [Eq. (\ref{eq:Qp1})]. The first term of the right-hand side corresponds to the housekeeping component given by Eq. (\ref{eq:Qhk}) and $Q^{\rm +SS}$ as shown in Fig. 3(b).  The second term corresponds to the excess component given by Eq. (\ref{eq:Qex_p1}). The sum of this excess component, i.e., $\sum_{N=0}^{\langle N_{\rm SS} \rangle-1}(\mu_{\rm eff} -\mu_{N+1})$, during the 'In' events is equal to $E_{\rm C}\langle N_{\rm SS} \rangle^2=\Delta F_{\rm SS}$. In contrast, since the work $-eV(t)$ is applied to the system, the remaining energy, $-eV(t)-Q_N^{+1}=\mu_{N+1}$, is stored in the dot as the free energy, and the sum of $\mu_N$ during the 'In' events becomes $\Delta F_{\rm SS}$. The agreement between these two values resembles the relationship between free energy and heat dissipation when charging a capacitor with a direct-constant (DC) voltage. In other words, the minimum energy required to accumulate free energy corresponds to the excess heat—the minimum energy required for the state transition.

In the 'In-Out' events,  the number of electrons entering and leaving the dot is balanced, and the net change $\Delta N$ in $N$ is zero. In this case, for each value of $N$, an electron entering the dot  dissipates $Q_N^{+1}=[-eV(t)-\mu_{\rm eff}] +(\mu_{\rm eff} -\mu_{N+1})$, while an electron leaving dissipates $Q_{N+1}^{-1}=[eV(t)+\mu_{\rm eff}] +(-\mu_{\rm eff} +\mu_{N+1})$. Here, the second terms correspond to the excess components and cancel between the 'In' and 'Out' processes. In contrast, the first terms represent the housekeeping contributions, which explicitly depend on $V(t)$ and therefore do not cancel at each instant. However, when taking the time average over the stochastic transitions, the contribution from $V(t)$ in $Q_{N+1}^{-1}$ vanishes because the time average of $V(t)$ is zero and the transition rates for electron leaving the dot are not correlated with $V(t)$ over a full period. Therefore, when we consider the events of entering and leaving the dot as a set , we can regard  $\langle \dot{Q}_{N}^{+1} \rangle$ and $\langle \dot{Q}_{N+1}^{-1} \rangle$ as $\langle \dot{Q}^{\rm +SS} \rangle $ and  $\langle \dot{Q}^{\rm -SS} \rangle$ given by Eqs. (\ref{eq:Qin_3}) and (\ref{eq:Qout_3}), respectively. It should be noted that these considerations are valid for the transient and steady states as shown in Fig. 3(c). 

Consequently, the total heat dissipation satisfies Eq. (\ref{eq:Qt+dF=Qhk}).  Here, the difference between the two terms corresponds to the free energy stored in the dot, 
\begin{align}
\Delta F_{\rm SS} = E_{\rm C}\langle N_{\rm SS}\rangle^2 
= \langle N_{\rm SS}\rangle \mu_{\rm eff} 
- \sum_{N=0}^{\langle N_{\rm SS}\rangle-1} (\mu_{\rm eff} - \mu_{N+1}),
\end{align}
which arises from the accumulated excess contributions during the 'In' events. Using $\langle \dot{Q}_{\rm T}(t) \rangle=\langle \dot{Q}_{\rm EX}(t) \rangle+\langle \dot{Q}_{\rm HK}(t) \rangle$  together with Eq. (\ref{eq:Qt+dF=Qhk}), 
we obtain an alternative expression for the free energy,
\begin{align}
\Delta F_{\rm SS}
= \frac{1}{2}\int_0^{\tau_{\rm SS}} 
\left[ \langle\dot{Q}_{\rm HK}^{\rm SS}\rangle - \langle\dot{Q}_{\rm HK}(t)\rangle \right] dt.
\end{align}
This relation shows that the free energy can be quantified from the deviation of the housekeeping heat from its steady-state value during the transient dynamics.

Finally, we examine the consistency of the above heat decomposition from a thermodynamic viewpoint. 
In a driven system, the heat dissipation must be balanced by the work continuously supplied to the system. Eq.~(\ref{eq:Qt_SS}) describes the work rate on the ER-dot system.\cite{Hofmann2017} Its time average is given by 
\begin{align}
\int_0^{2\pi/\omega_{\rm AC}}[-eV(t)\varGamma(t)^+]dt=2\gamma E_{\rm C}\langle N_{\rm SS}\rangle \overline{\varGamma^+},\label{eq:work_rate}
 \end{align}
where $\gamma$ monotonically decreases with an increase in $\beta S_{\rm AC}$. Specifically, $\gamma$ approximates 1 if $\beta eS_{\rm AC} \gg 1$, and $\gamma$ approximates 2 if $\beta eS_{\rm AC} \ll 1$ [see Appendix \ref{ap:HK}]. This result confirms that the continuous energy input from the AC drive accounts for the sustained housekeeping heat dissipation in both transient and steady regimes.

\section{\label{sec:efficiency} Efficiency and its limit}

Having established the decomposition of heat dissipation and the associated energy balance, we now turn to the efficiency of energy conversion in the driven ER-dot system, based on the second law for excess heat and the energy balance established above.

\begin{figure}[t]
\includegraphics{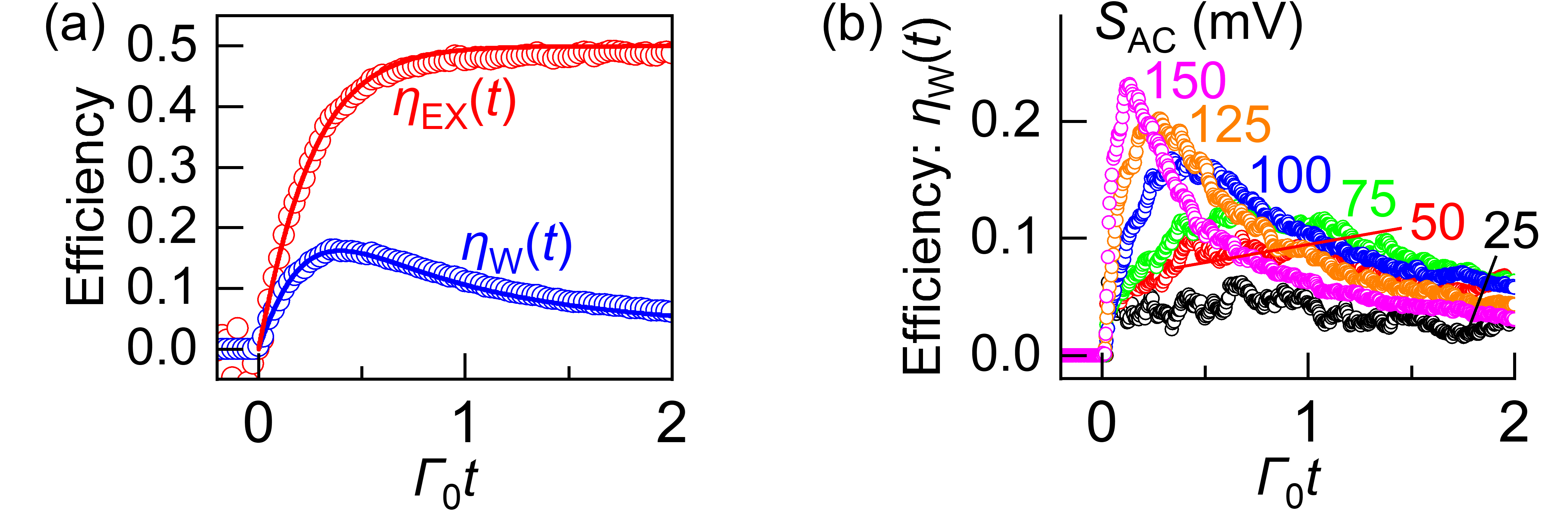}% Here is how to import EPS art
\caption{\justifying\label{Figure 4} (a) Change in efficiencies  $\eta_{\rm EX}$ and $\eta_{\rm W}$. $S_{\rm AC}=100$ mV. Open circles are  experimental results averaged over 30000 repetitions. The solid lines are theoretical results obtained from the master equation. (b) $S_{\rm AC}$ dependencies of $\eta_{\rm W}$.  Each curve is averaged over 250 repetitions. }\end{figure}

The correlation between $\langle\dot{Q}_{\rm EX}(t)\rangle$ and $\Delta F(t)$ provides a thermodynamic basis for defining the efficiency $\eta_{\mathrm{EX}}(t)$ through the second law for excess heat. Hatano and Sasa \cite{Hatano2001} showed $
\langle Q_{\mathrm{EX}}(t) \rangle + k_B T \Delta S(t) \geq 0.$ Using the non-equilibrium free energy relation $\Delta F(t) = \Delta U(t) - k_B T \Delta S(t)$, we obtain $\langle Q_{\mathrm{EX}}(t) \rangle + \Delta U(t) \geq \Delta F(t)$. This inequality naturally suggests defining an efficiency that quantifies how much of the total energy input is converted into free energy: 
\begin{equation}
\eta_{\mathrm{EX}}(t) = \frac{\Delta F(t)}{\langle Q_{\mathrm{EX}}(t) \rangle + \Delta U(t)}. \label{eq:Efficiency_EX}
   \end{equation}
As shown in Fig.~\ref{Figure 4}(a), $\eta_{\mathrm{EX}}(t)$ increases and then converges to a steady-state value. Since $\Delta F(t) \approx \Delta U(t)$ and $\langle Q_{\mathrm{EX}}^{\mathrm{SS}} \rangle = \Delta F_{\mathrm{SS}}$, the upper bound of $\eta_{\rm EX}(t)$ is 0.5 in the NESS. This result implies that, in the NESS, at most half of the input energy can be converted into free energy, while the remaining half is inevitably dissipated. Although the microscopic mechanisms differ, this balance is reminiscent of classical capacitor charging, where half of the supplied energy is stored and half is inevitably dissipated as explained above [see Fig. 3(c)]. 

These understandings based on Fig. \ref{Figure 3}(c) also offer information on the energetic efficiency for free-energy generation:
\begin{equation}
    \eta_{\rm W}(t)=\frac{\Delta F(t)}{[\langle Q_{\rm T}(t)\rangle +\Delta U(t)]}, \label{eq:Efficiency_W}
\end{equation}
\cite{Datta2022} where $ \langle Q_{\rm T}(t)\rangle=\int_0^t \langle\dot{Q}_{\rm T}(t')\rangle dt'$. The denominator corresponds to the total input work, as it includes both the accumulated heat dissipation and the stored internal energy. Figure \ref{Figure 4}(a) shows the transient characteristics of $\eta_{\rm W}(t)$: $\eta_{\rm W}(t)$ rises, peaks, and then gradually decreases over time unlike $\eta_{\rm EX}(t)$, which monotonically approaches 0.5. Additionally, increasing $S_{\rm AC}$ raises the peak values of $\eta_{\rm W}(t)$, which reach up to 0.25 at 150 mV in experiments as shown in Fig. \ref{Figure 4}(b).

To characterize $\eta_{\rm W}(t)$, we consider the efficiency $\eta_{\rm SS}$ when the system reaches the NESS at $t=\tau_{\rm SS}$. Since $\Delta F(\tau_{\rm SS})$ is its maximum $\Delta F_{\rm SS}$ given by $E_{\rm C}\langle N_{\rm SS}\rangle^2$ and nearly equal to $\Delta U(\tau_{\rm SS})$, Eq. (\ref{eq:Qt+dF=Qhk}) gives 
\begin{equation}
    \eta_{\rm  SS} = \frac{E_{\rm C}\langle N_{\rm SS}\rangle^2}{\int_0^{\tau_{\rm SS}}\langle\dot{Q}_{\rm HK}^{\rm SS}\rangle dt}.
    \label{eq:eta}
\end{equation}
We define $\tau_{\rm th}$ as the moment when $\langle N(t)\rangle$ reaches a threshold value of $a\langle N_{\rm SS}\rangle$, where $0<a<1$ [see Appendix \ref{ap:TransChar}]. Since $\langle\dot{Q}_{\rm HK}^{\rm SS}\rangle=2\gamma E_{\rm C}\langle N_{\rm SS}\rangle \overline{\varGamma^+}$ as explained above, $\eta_{\rm SS}$ is given by  
\begin{align}
    \eta_{\rm SS} &\sim -\frac{(\beta eS_{\rm AC})^2}{16\ln(1-a)} \quad \text{ if } \beta eS_{\rm AC} \ll 1, \text{ or} \label{eq:eta_S}\\
    \eta_{\rm SS} &\sim \frac{1}{a+1} = 0.5 \qquad \text{ if } \beta eS_{\rm AC} \gg 1  \text{ and } a=1.\label{eq:eta_L}
\end{align}
The independence of Eqs. (\ref{eq:eta_S}) and (\ref{eq:eta_L}) from $E_{\rm C}$ highlights a universal scaling behavior: the maximum efficiency is determined solely by the driving strength $\beta S_{\rm AC}$, which quantifies how far the system is driven from equilibrium. 
When $\beta S_{\rm AC} \gg 1$, the efficiency $\eta_{\rm SS}$ approaches its upper bound of 0.5 in the NESS. This trend is counterintuitive: a larger $S_{\rm AC}$ drives the system further from equilibrium, and one would normally expect stronger irreversibility to reduce efficiency.  This seemingly anomalous increase in efficiency arises because a larger $S_{\rm AC}$ substantially suppresses the housekeeping heat, as illustrated in Fig.~\ref{Figure 3}(a).  An increase in $S_{\rm AC}$ reduces the parameter $\gamma$, making $2\gamma E_{\rm C}\langle N_{\rm SS}\rangle$ close to $\mu_{\rm eff}$, which in turn decreases the housekeeping component of $\langle\dot{Q}^{\rm +SS}(t)\rangle$. Simultaneously, a larger $S_{\rm AC}$ shortens the approach time $\tau_{\rm th}$ [see Eq.~(\ref{eq:t_th_large}) in Appendix~\ref{ap:TransChar}], thereby reducing $\langle\dot{Q}^{\rm +}(t)\rangle$ and $\langle\dot{Q}^{\rm -SS}(t)\rangle$, i.e.,  $\langle Q_{\rm HK}(t)\rangle$, before the NESS is reached.   In other words, $\langle Q_{\rm T}(t)\rangle \approx \langle Q_{\rm EX}(t)\rangle$ until the NESS, implying that $\eta_{\rm W}(t)$ becomes nearly identical to $\eta_{\rm EX}(t)$.  In contrast, when $\beta S_{\rm AC} \ll 1$, the system approaches the NESS slowly (large $\tau_{\rm SS}$) and $\gamma$ increases toward $2$, enhancing $\langle Q_{\rm HK}(t)\rangle$ and thereby lowering the efficiency.

\section{\label{sec:summary} Summary}
We experimentally decomposed the heat dissipation of a multi-electron dot into excess and housekeeping components and revealed that free-energy generation is fundamentally limited by their competition. Our analysis shows that the efficiency can approach the upper bound of 0.5 under strong non-equilibrium driving, independent of device-specific parameters. By introducing an effective time-averaged transition rate for AC-driven dynamics, we obtain a compact description of periodically driven stochastic electron systems and a quantitative thermodynamic framework linking heat decomposition to free-energy generation. These results enable systematic evaluation of performance limits in electronic devices operated far from equilibrium.   
 
\begin{acknowledgments}
The authors thank Yasuhiko Tokura in Tsukuba universty for valuable support on the theory. They also thank Keiji Saito and Tan Van Vu in Kyoto university for fruitful discussion. 
\end{acknowledgments}

\appendix

\section{Solution of master equation} \label{ap:MEsolution}
When an AC signal is applied, the ER-dot system is described by the master equation given by Eq. (\ref{eq:ME}) . Since this AC signal oscillating  with a frequency of $\omega_{\rm AC}/2\pi$ is estimated to bring the periodicity to $\varGamma^+(t)$ and $\rho_N(t)$, we applied the Fourier series expansion to them, respectively:
\begin{align}
       &\varGamma^{+(F)}(t)/\varGamma_0=\exp[A\cos(\omega_{\rm AC}t)]\notag\\
       &\hspace{15mm}\sim I_0(A)+2 \sum_{k=1}^K I_k(A)\cos(k\omega_{\rm AC}t) \text{ and} \label{eq:Gamma_Fourier}\\
       &\rho^F_N(t)=\frac{p_0^{(N)}(t)}{2}\notag\\
       &\hspace{2mm}+\sum_{l=1}^K \left[ p_l^{(N)}(t)\cos(l\omega_{\rm AC}t)+q_l^{(N)}(t)\sin(l\omega_{\rm AC}t)\right].   \label{eq:Prob_Fourier}
\end{align}
$I_k(A)=2/T_{\rm AC} \int \exp[A\cos(\omega_{\rm AC}t)]\cos(2k\pi/T_{\rm AC})dt$ is the modified Bessel function of the first kind of order $k$, where $A=\beta eS_{\rm AC}$ and $T_{\rm AC}=2\pi/\omega_{\rm AC}$. $K$ of 12 makes $\varGamma^{+(F)}(t)$ sufficiently close to $\varGamma^+(t)$. By substituting Eqs. (\ref{eq:Gamma_Fourier}) and (\ref{eq:Prob_Fourier}) into Eq. (\ref{eq:ME}),  $\rho^F_N(t)$ is derived numerically. 

When $\omega_{\rm AC}/2\pi \ll \varGamma_0$, instantaneous distribution of $\rho^F_N(t)$ changes over time, keeping Gaussian distribution, and its average $\langle N(t) \rangle=\Sigma_N N\rho^F_N(t)$ changes periodically following the AC signal [Figs. A1(a) and (b)]. These features are natural because the ER voltage changes slowly enough for the system to follow it quasi-statically. 

\begin{figure}[b]
\includegraphics[width=\columnwidth]{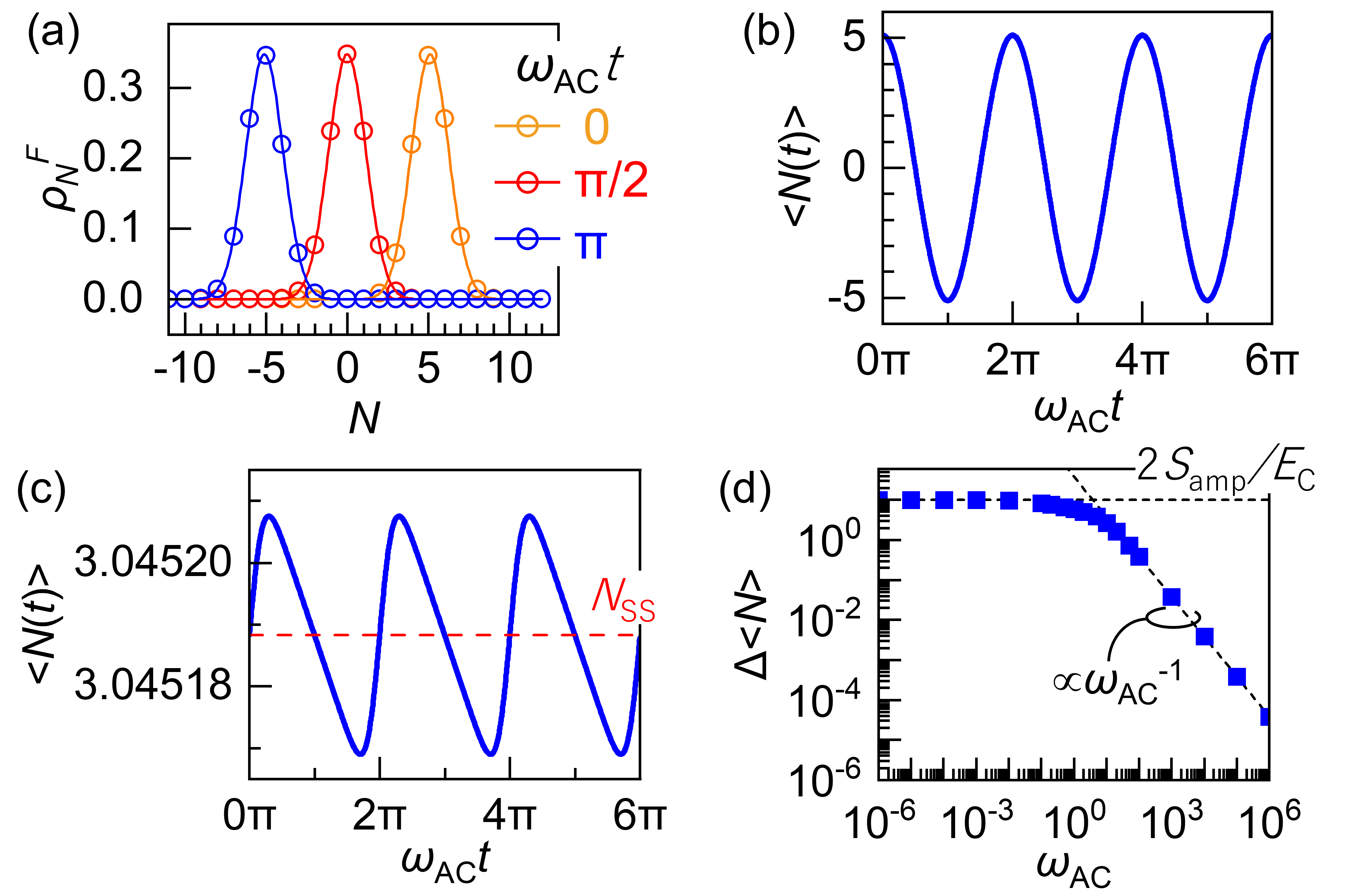}
\captionsetup{labelformat=empty}
\caption{\justifying \label{Fig-appendix_NESS}FIG. A1. (a) $\rho_N^F(t)$ at different time in the NESS. Open circles are given by Eq. (\ref{eq:ME}) into which Eqs. (\ref{eq:Gamma_Fourier}) and (\ref{eq:Prob_Fourier}) are substituted. The solid lines are Gaussian curves fitted to the open circles. Change in $\langle N(t) \rangle$ when (b) $\omega_{\rm AC}/2\pi=10^{-6}\varGamma_0$ and (c) $\omega_{\rm AC}/2\pi=10^6 \varGamma_0$. (d) $\omega_{\rm AC}$ dependence of $\Delta \langle N \rangle$, which is a peak-to-peak value of $\langle N(t) \rangle$ over time. }
\end{figure}
When $\omega_{\rm AC}/2\pi \gg \varGamma_0$ in a NESS, temporal change in $\rho^F_N(t)$ is too small to be distinguishable as shown in Fig. A1(c) depicting that $\langle N(t) \rangle$ fluctuates by about 10 ppm. This fluctuation decreases inversely proportional to $\omega_{\rm AC}$ when $\omega_{\rm AC}/2\pi \gg \varGamma_0$ as shown in Fig. A1(d). It should be noted that time average of this $\langle N(t) \rangle$ is the same as $\langle N_{\rm SS} \rangle=\ln[I_0(A)]/2\beta E_{\rm C}$ [also see Supplemental Material]. The variance of $\rho^F_N(t)$ is also approximated to $(2\beta E_{\rm C})^{-1}$. Additionally, the data is averaged with random phase of the AC signal in our experiments. These features imply that when an AC signal with is applied, $\rho^F_N(t)$ approaches to $\rho_N^{\rm SS}=Z^{-1} \exp[-\beta E_{\rm C} (N-\langle N_{\rm SS} \rangle)^2]$. 

Next, we consider transient characteristics from the equilibrium to the NESS. After the AC signal is applied, $\langle N(t) \rangle$ given by $\Sigma_N N\rho^F_N(t)$ increases with periodic oscillations, while $\langle N(t) \rangle$ given by $\Sigma_N N p_0^{(N)}(t)$ increases monotonically [Fig. A2(a)]. The discrepancy between them is small enough so that the transient characteristics can be represented by  $\Sigma_N N p_0^{(N)}(t)$, i.e. DC part of the Fourier series expansion of Eq. (\ref{eq:Prob_Fourier}). 

These features after and before the NESS imply that probability density can be represented by the DC part of the Fourier series expansion and thus that $\varGamma^+$ is represented by the DC part of  $\varGamma^{+(F)}$, i.e., $\varGamma_0I_0(A)=\overline{\varGamma^+}$.  When we solve the master equation given by Eq. (\ref{eq:ME}) in the assumption that $\varGamma_0$ is changed to $\overline{\varGamma^+}$ at $t=0$, the derived transient characteristics of $\langle N(t) \rangle$ show good agreement with those of $\Sigma_N N p_0^{(N)}(t)$ as shown in Fig. A2(b), which enables us to use $\overline{\varGamma^+}$ for the theoretical analysis of the master equation in the main text.

Considering transient characteristics of the probability distribution $\rho_N(t)$, $\Delta F(t)$ can be also approximated. $U(t)$ is given by $\Sigma_N E_C N^2 \rho_N(t)$. Since $\rho_N(t)$ has skewness and kurtosis close to zero and three, respectively, as shown in Fig. 2(c), it can be approximated to a Gaussian distribution given by $Z^{-1} \exp[-(N-\langle N(t)\rangle)^2/2{\rm Var}(t)]$, where ${\rm Var}(t)$ is time-dependent variance of $\rho_N(t)$. This approximation gives $U(t)=E_C\langle N(t)\rangle ^2+E_C{\rm Var}(t)$ and  
\begin{equation}
\begin{split}
    \Delta U(t)&=U(t)-U(0)\\
    &=E_C\langle N(t)\rangle ^2+\frac{k_BT}{2}\left[\frac{{\rm Var}(t)}{{\rm Var}(0)}-1\right].
   \label{eq:dU}   
\end{split}
\end{equation}
Here, we use $\langle N(0)\rangle=0$. Similarly, $S(t)=\Sigma_N\rho_N(t) \ln \rho_N(t)$ gives
\begin{equation}
\begin{split}
    \Delta S(t)=S(t)-S(0)=\frac{1}{2}[\ln{\rm Var}(t)-\ln{\rm Var}(0)].   \label{eq:dS}   
\end{split}
\end{equation}
Using Eqs. (\ref{eq:dU}) and (\ref{eq:dS}) gives
%\begin{widetext}
\begin{align}
    &\Delta F(t)=\Delta U(t)-k_BT\Delta S(t) \notag\\
   &\quad=E_C \langle N(t) \rangle^2 +\frac{k_BT}{2} \left[ \frac{{\rm Var}(t)}{{\rm Var}(0)}-1-\ln \frac{{\rm Var}(t)}{{\rm Var}(0)} \right]\notag\\
    &\quad \sim E_C \langle N(t) \rangle^2+\frac{k_BT}{2} \left[\frac{1}{2}X(t)^2+\mathcal{O}\left[X(t)\right]^3\right],  \label{eq:dF_simple}   
\end{align}
%\end{widetext}
where $X(t)={\rm Var}(t)/{\rm Var}(0)-1$. The second term proportional to $k_BT$ is significantly smaller than the leading term $E_C \langle N(t) \rangle^2$. Consequently, the free energy difference $\Delta F(t)$ can be approximated as $\Delta F(t) \approx E_C \langle N(t) \rangle^2$.

\begin{figure}[t]
\includegraphics[width=\columnwidth]{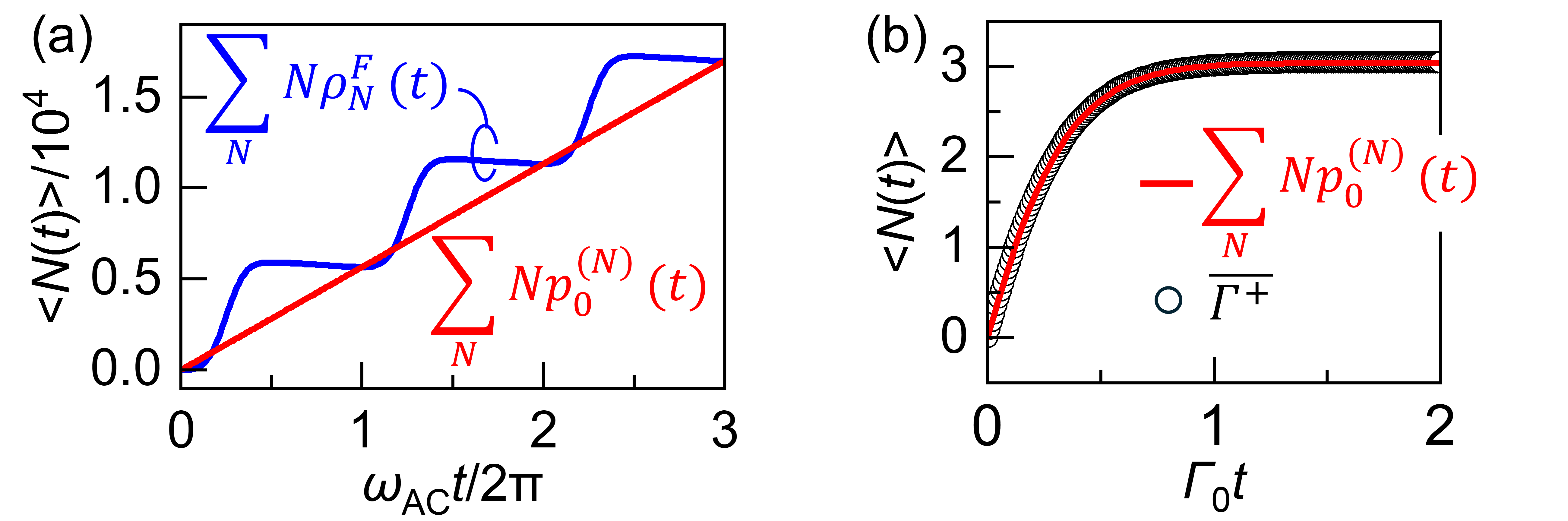}
\captionsetup{labelformat=empty}
\caption{\justifying \label{Fig-appendix_transient}FIG. A2. Change in $\langle N(t) \rangle$ before the NESS. $\omega_{\rm AC}/2\pi=10^6\varGamma_0$ Open circles in (b) show transient characteristics in the simplified model in the assumption that the transition rate for electrons entering the dot changes from $\varGamma_0$ to $\overline{\varGamma^+}$ at $t=0$.}
\end{figure}

\section{Transient characteristics until a NESS} \label{ap:TransChar}

The transient characteristics of $ \langle N(t) \rangle$ is given by \cite{Salhani2024}
\begin{equation}
\begin{split}
    &\langle N(t) \rangle=\frac{\langle N_{\rm SS}\rangle}{\ln(\overline{\varGamma^+}/\varGamma_0)} \\
    & \qquad \times \ln \left[ \frac{\overline{\varGamma^+}/\varGamma_0}{1+(\overline{\varGamma^+}/\varGamma_0)-1) \exp(-2\beta E_{\rm C}\overline{\varGamma^+} t)}
    \right]. \label{eqn:N(t)}
\end{split}
\end{equation}
The moment $\tau_{\rm th}$ when $\langle N(t) \rangle$  reaches a threshold value of $a\langle N_{\rm SS}\rangle$, where $0<a<1$, is given by
\begin{equation}
    \tau_{\rm th}= \frac{\langle N_{\rm SS} \rangle}{\overline{\varGamma^+}} \frac{1}{\ln(\overline{\varGamma^+}/\varGamma_0)} \ln \left[ \frac{\overline{\varGamma^+}/\varGamma_0-1}{(\overline{\varGamma^+}/\varGamma_0)^{1-a}-1}
    \right].\label{eq:t_th}
    \end{equation}
When $1/(1-a) \in N$ 
\begin{equation}
    \ln \left[ \frac{\overline{\varGamma^+}/\varGamma_0-1}{(\overline{\varGamma^+}/\varGamma_0)^{1-a}-1}   \right]=\ln \left[\sum_{i=0}^{\frac{1}{1-a}-1}\left(\frac{\overline{\varGamma^+}}{\varGamma_0} \right)^{(1-a)i} \right]. \label{eq:Y}
\end{equation}
From Appendix \ref{ap:MEsolution},
\begin{equation}
\begin{split}
 \overline{\varGamma^+}&=\frac{\omega_{\rm AC}}{2\pi}\int_0^{2\pi/\omega_{\rm AC}} \varGamma_0\exp[\beta eS_{\rm AC}\sin(\omega_{\rm AC}t)]dt \\
 &=\varGamma_0 I_0(\beta eS_{\rm AC}), \label{eq:G_AC_I0}
\end{split}
\end{equation}
where 
\begin{equation}
I_0(x)=\sum_{s=0}^\infty \frac{1}{s!s!}\left(\frac{x}{2}\right)^{2s}.
\end{equation}
When $\beta S_{\rm AC} \ll 1$, Eq. (\ref{eq:G_AC_I0}) is simplified to
\begin{equation}
    \frac{\overline{\varGamma^+}}{\varGamma_0} \sim 1+\frac{1}{4}(\beta eS_{\rm AC})^2. \label{eq:G_AC_small}
\end{equation}
%Using the Taylor expansion and Eq. (\ref{eq:G_AC_small}),
%\begin{equation}
%    \ln \left(\frac{\overline{\varGamma^+}}{\varGamma_0} \right) \sim \frac{1}%{4}(\beta eS_{\rm AC})^2. \label{eq:LN_G_AC_small}
%\end{equation}
Using Eqs. (\ref{eq:Y}), (\ref{eq:G_AC_small}), and the Taylor expansion, Eqs. (\ref{eq:t_th}) and (\ref{eqn:N(t)}) can be converted to
\begin{align}
    &\tau_{\rm th}\sim\frac{\langle N_{\rm SS} \rangle}{\overline{\varGamma^+}} \frac{4}{(\beta eS_{\rm AC})^2} \ln \left(\frac{1}{1-a}\right) \text{ and}\label{eq:t_th_small}\\
    & \langle N(t) \rangle \sim \frac{\langle N_{\rm SS}\rangle(\beta eS_{\rm AC})^2}{4\ln(\overline{\varGamma^+}/\varGamma_0)}[1-\exp(-2\beta E_{\rm C}\overline{\varGamma^+} t)], \label{eq:Nt_small}
\end{align}
respectively. Eq. (\ref{eq:Nt_small}) represents the transient characteristics with a time constant of $(2\beta E_{\rm C}\overline{\varGamma^+})^{-1}$. 

When $\beta S_{\rm AC} \gg 1$, the asymptotic form of the Bessel functions simplifies Eq. (\ref{eq:G_AC_I0}) to
\begin{align}
     &\frac{\overline{\varGamma^+}}{\varGamma_0}=\frac{\exp(\beta eS_{\rm AC})}{\sqrt{2\pi\beta eS_{\rm AC}}} \left\{ 1+\frac{1}{8\beta eS_{\rm AC}}+\mathcal{O}\left[ \frac{1}{(\beta S_{\rm AC})^2}\right] \right\} \notag \\
     &\qquad \sim \frac{\exp(\beta eS_{\rm AC})}{\sqrt{2\pi\beta eS_{\rm AC}}}.\label{eq:G_AC_large}
\end{align}
Using Eqs. (\ref{eq:t_th}), (\ref{eq:Y}), (\ref{eq:G_AC_large}) and $\ln (\overline{\varGamma^+}/\varGamma_0) \sim \beta eS_{\rm AC}$,
\begin{equation}
    \tau_{\rm th}\sim\frac{\langle N_{\rm SS} \rangle}{\overline{\varGamma^+}}a. \label{eq:t_th_large}
\end{equation}

\section{Housekeeping heat in a NESS}\label{ap:HK}

Since $\langle\dot{Q}_{\rm HK}^{\rm SS}\rangle$ is equal to $\langle\dot{Q}_{\rm T}^{\rm SS}\rangle$, shown in Eq. (\ref{eq:Qt_SS}), in a NESS, its time-average can be given by 
\begin{equation}
\begin{split}
\langle\dot{Q}_{\rm HK}^{\rm SS}\rangle&=-\frac{\omega_{\rm AC}}{2\pi}\int_0^{2\pi/\omega_{\rm AC}}eS_{\rm AC} \sin(\omega_{\rm AC}t) \\
&\hspace{20mm}\times\varGamma_0\exp[-eS_{\rm AC}\sin(\omega_{\rm AC}t)]dt\\
    &=\varGamma_0 eS_{\rm AC}I_1(\beta eS_{\rm AC}), \label{eq:Qhk_SS}
\end{split}
\end{equation}
where 
\begin{equation}
\begin{split}
I_1(x)=\sum_{s=0}^\infty \frac{1}{s!(s+1)!}\left(\frac{x}{2}\right)^{(2s+1)}
\end{split}
\end{equation}
is the integral formula for the modified Bessel function. When $\beta eS_{\rm AC} \ll 1$, Eqs. (\ref{eq:G_AC_small}) and the Taylor series expansion approximate Eq. (\ref{eq:Qhk_SS}):
\begin{equation}
\begin{split}
\langle\dot{Q}_{\rm HK}^{\rm SS}\rangle &\sim  \varGamma_0 eS_{\rm AC} \left[\frac{1}{2}(\beta eS_{\rm AC})+\frac{1}{16}(\beta eS_{\rm AC})^3 \right]\\
&= \varGamma_0 \frac{(\beta eS_{\rm AC})^2}{2\beta}\left[1+\frac{1}{8}(\beta eS_{\rm AC})^2 \right]\\
&\sim 4E_{\rm C}\langle N_{\rm SS}\rangle \overline{\varGamma^+} \label{eq:Qhk_SS_small}
\end{split}
\end{equation}
When $\beta eS_{\rm AC} \gg 1$, the asymptotic form of the modified Bessel functions, Eq. (\ref{eq:G_AC_large}), and $\ln (\overline{\varGamma^+}/\varGamma_0) \sim \beta eS_{\rm AC}$ simplify Eq. (\ref{eq:Qhk_SS}) to
\begin{equation}
\begin{split}
\langle\dot{Q}_{\rm HK}^{\rm SS}\rangle \sim \varGamma_0 eS_{\rm AC} \frac{\exp(\beta eS_{\rm AC})}{\sqrt{2\pi\beta eS_{\rm AC}}}\sim 2E_{\rm C}\langle N_{\rm SS}\rangle \overline{\varGamma^+}. \label{eq:Qhk_SS_large}
\end{split}
\end{equation}

\bibliography{apssamp}% Produces the bibliography via BibTeX.

@article{Ando2024,
  title = {Nonequilibrium heat dissipation as a probe for detecting wavefront distortion in microscopy},
  author = {Ando, Taro and Otsu-Hyodo, Tomoko},
  journal = {Phys. Rev. Res.},
  volume = {6},
  issue = {4},
  pages = {043263},
  numpages = {7},
  year = {2024},
  month = {Dec},
  publisher = {American Physical Society},
  doi = {10.1103/PhysRevResearch.6.043263},
  url = {https://link.aps.org/doi/10.1103/PhysRevResearch.6.043263}
}

@article{Averin2011,
   author = {Averin, D. V. and Pekola, J. P.},
   title = {Statistics of the dissipated energy in driven single-electron transitions},
   journal = {EPL (Europhysics Letters)},
   year = {2011},
   volume = {96},
   pages = {67004},
   number = {6},
   DOI = {10.1209/0295-5075/96/67004},
}

@article{Barker2022,
   author = {Barker, D. and Scandi, M. and Lehmann, S. and Thelander, C. and Dick, K. A. and Perarnau-Llobet, M. and Maisi, V. F.},
   title = {Experimental Verification of the Work Fluctuation-Dissipation Relation},
   journal = {Phys. Rev. Lett.},
   year = {2022},
   volume = {128},
   number = {4},
   pages = {040602},
   DOI = {10.1103/PhysRevLett.128.040602},
}

@article{Berut2012,
   author = {Bérut, A. and Arakelyan, A. and Petrosyan, A. and Ciliberto, S. and Dillenschneider, R. and Lutz, E.},
   title = {Experimental verification of Landauer's principle linking information and thermodynamics},
   journal = {Nature},
   year = {2012},
   volume = {483},
   pages = {187-189},
   number = {7388},
   DOI = {10.1038/nature10872},
}

@article{Berut2015,
doi = {10.1088/1742-5468/2015/06/P06015},
url = {https://dx.doi.org/10.1088/1742-5468/2015/06/P06015},
year = {2015},
month = {jun},
publisher = {IOP Publishing and SISSA},
volume = {2015},
number = {6},
pages = {P06015},
author = {Bérut, Antoine and Petrosyan, Artyom and Ciliberto, Sergio},
title = {Information and thermodynamics: experimental verification of Landauer's Erasure principle},
journal = {J. Stat. Mech.},

}

@article{Broeck2010,
  title = {Three faces of the second law. II. Fokker-Planck formulation},
  author = {Van den Broeck, Christian and Esposito, Massimiliano},
  journal = {Phys. Rev. E},
  volume = {82},
  issue = {1},
  pages = {011144},
  numpages = {7},
  year = {2010},
  month = {Jul},
  publisher = {American Physical Society},
  doi = {10.1103/PhysRevE.82.011144},
  url = {https://link.aps.org/doi/10.1103/PhysRevE.82.011144}
}

@article{Carles2015,
doi = {10.7567/JJAP.54.06FG03},
url = {https://dx.doi.org/10.7567/JJAP.54.06FG03},
year = {2015},
month = {apr},
publisher = {The Japan Society of Applied Physics},
volume = {54},
number = {6S1},
pages = {06FG03},
author = {Pierre-Alix Carles and Katsuhiko Nishiguchi and Akira Fujiwara},
title = {Deviation from the law of energy equipartition in a small dynamic-random-access memory},
journal = {Jpn. J. Appl. Phys.},

}

@article{Datta2022,
  title = {Second Law for Active Heat Engines},
  author = {Datta, Arya and Pietzonka, Patrick and Barato, Andre C.},
  journal = {Phys. Rev. X},
  volume = {12},
  issue = {3},
  pages = {031034},
  numpages = {19},
  year = {2022},
  month = {Sep},
  publisher = {American Physical Society},
  doi = {10.1103/PhysRevX.12.031034},
  url = {https://link.aps.org/doi/10.1103/PhysRevX.12.031034}
}

@article{Esposito2007,
   author = {Esposito, Massimiliano and Harbola, Upendra and Mukamel, Shaul},
   title = {Entropy fluctuation theorems in driven open systems: Application to electron counting statistics},
   journal = {Phys. Rev. E},
   year = {2007},
   volume = {76},
   pages = {031132},
   number = {3},
   DOI = {10.1103/PhysRevE.76.031132},
}

@article{Esposito2010,
   author = {Esposito, Massimiliano and Van den Broeck, Christian},
   title = {Three faces of the second law. {I.} {Master} equation formulation},
   journal = {Phys. Rev. E},
   year = {2010},
   volume = {82},
   pages = {011143},
   number = {1},
   DOI = {10.1103/PhysRevE.82.011143},
}

@article{Esposito2011,
doi = {10.1209/0295-5075/95/40004},
url = {https://doi.org/10.1209/0295-5075/95/40004},
year = {2011},
month = {aug},
publisher = {},
volume = {95},
number = {4},
pages = {40004},
author = {Esposito, M. and Van den Broeck, C.},
title = {Second law and Landauer principle far from equilibrium},
journal = {Europhys. Lett.}
}

@article{Gaspard2004,
    author = {Gaspard, Pierre},
    title = "{Fluctuation theorem for nonequilibrium reactions}",
    journal = {The Journal of Chemical Physics},
    volume = {120},
    number = {19},
    pages = {8898-8905},
    year = {2004},
    month = {05},
    issn = {0021-9606},
    doi = {10.1063/1.1688758},
}

@article{Hatano2001,
   author = {Hatano, T. and Sasa, S.},
   title = {Steady-state thermodynamics of Langevin systems},
   journal = {Phys Rev Lett},
   year = {2001},
   volume = {86},
   pages = {3463-3466},
   number = {16},
   DOI = {DOI 10.1103/PhysRevLett.86.3463},
}

@article{Hofmann2017,
   author = {Hofmann, A. and Maisi, V. F. and Basset, J. and Reichl, C. and Wegscheider, W. and Ihn, T. and Ensslin, K. and Jarzynski, C.},
   title = {Heat dissipation and fluctuations in a driven quantum dot},
   journal = {Phys. Status. Solidi. B},
   year = {2017},
   volume = {254},
   number = {3},
   pages = {1600546},
   DOI = {10.1002/pssb.201600546},
}

@article{Hofmann2016,
   author = {Hofmann, A. and Maisi, V. F. and Rössler, C. and Basset, J. and Krähenmann, T. and Märki, P. and Ihn, T. and Ensslin, K. and Reichl, C. and Wegscheider, W.},
   title = {Equilibrium free energy measurement of a confined electron driven out of equilibrium},
   journal = {Phys. Rev. B},
   year = {2016},
   volume = {93},
   pages = {035425},
   number = {3},
   DOI = {10.1103/PhysRevB.93.035425},
}

@article{Hong2016,
author = {Jeongmin Hong  and Brian Lambson  and Scott Dhuey  and Jeffrey Bokor },
title = {Experimental test of Landauer’s principle in single-bit operations on nanomagnetic memory bits},
journal = {Science Advances},
volume = {2},
number = {3},
pages = {e1501492},
year = {2016},
doi = {10.1126/sciadv.1501492},
URL = {https://www.science.org/doi/abs/10.1126/sciadv.1501492},
eprint = {https://www.science.org/doi/pdf/10.1126/sciadv.1501492},
}

@article{Jun2014,
  title = {High-Precision Test of Landauer's Principle in a Feedback Trap},
  author = {Jun, Yonggun and Gavrilov, Mom\ifmmode \check{c}\else \v{c}\fi{}ilo and Bechhoefer, John},
  journal = {Phys. Rev. Lett.},
  volume = {113},
  issue = {19},
  pages = {190601},
  numpages = {5},
  year = {2014},
  month = {Nov},
  publisher = {American Physical Society},
  doi = {10.1103/PhysRevLett.113.190601},
  url = {https://link.aps.org/doi/10.1103/PhysRevLett.113.190601}
}

@article{Koski2014_szilard,
   author = {Koski, J. V. and Maisi, V. F. and Pekola, J. P. and Averin, D. V.},
   title = {Experimental realization of a Szilard engine with a single electron},
   journal = { Proc. Natl. Acad. Sci. USA},
   year = {2014},
   volume = {111},
   pages = {13786-9},
   number = {38},
   DOI = {10.1073/pnas.1406966111},
}

@article{Kramers1940,
   author = {Kramers, H. A.},
   title = {Brownian motion in a field of force and the diffusion model of chemical reactions},
   journal = {Physica},
   year = {1940},
   volume = {7},
   pages = {284-304},
   number = {4},
   DOI = {10.1016/s0031-8914(40)90098-2},
}

@article{Kung2012,
   author = {Küng, B. and Rössler, C. and Beck, M. and Marthaler, M. and Golubev, D. S. and Utsumi, Y. and Ihn, T. and Ensslin, K.},
   title = {Irreversibility on the Level of Single-Electron Tunneling},
   journal = {Phys. Rev. X},
   year = {2012},
   volume = {2},
   pages = {011001},
   number = {1},
   DOI = {10.1103/PhysRevX.2.011001},
}

@article{Landauer1961,
   author = {Landauer, R.},
   title = {Irreversibility and heat generation in the computing process},
   journal = {IBM J. Res. Dev.},
   year = {1961},
   volume = {5},
   pages = {183-91},
   number = {3},
   DOI = {10.1147/rd.53.0183},
}

@article{Lee2022,
   author = {Lee, J. S. and Lee, S. Y. and Kwon, H. and Park, H.},
   title = {Speed Limit for a Highly Irreversible Process and Tight Finite-Time Landauer's Bound},
   journal = {Phys. Rev. Lett.},
   year = {2022},
   volume = {129},
   pages = {120603},
   number = {12},
   DOI = {10.1103/PhysRevLett.129.120603},
}

@article{Martini2016,
title = {Experimental and theoretical analysis of Landauer erasure in nano-magnetic switches of different sizes},
journal = {Nano Energy},
volume = {19},
pages = {108-116},
year = {2016},
issn = {2211-2855},
doi = {https://doi.org/10.1016/j.nanoen.2015.10.028},
url = {https://www.sciencedirect.com/science/article/pii/S2211285515004073},
author = {L. Martini and M. Pancaldi and M. Madami and P. Vavassori and G. Gubbiotti and S. Tacchi and F. Hartmann and M. Emmerling and S. Höfling and L. Worschech and G. Carlotti},
}

@article{Mounier2012,
doi = {10.1209/0295-5075/100/30002},
url = {https://doi.org/10.1209/0295-5075/100/30002},
year = {2012},
month = {nov},
publisher = {EDP Sciences, IOP Publishing and Società Italiana di Fisica},
volume = {100},
number = {3},
pages = {30002},
author = {Mounier, Anne and Naert, Antoine},
title = {The Hatano-Sasa equality: Transitions between steady states in a granular gas},
journal = {Europhysics Letters},
}

@article{Nishiguchi2008JJAP,
doi = {10.1143/JJAP.47.8305},
url = {https://dx.doi.org/10.1143/JJAP.47.8305},
year = {2008},
month = {nov},
publisher = {},
volume = {47},
number = {11R},
pages = {8305},
author = {Katsuhiko Nishiguchi and Charlie Koechlin and Yukinori Ono and Akira Fujiwara and Hiroshi Inokawa and Hiroshi Yamaguchi},
title = {Single-Electron-Resolution Electrometer Based on Field-Effect Transistor},
journal = {Jpn. J. Appl. Phys.},
}

@article{Nishiguchi2011APL,
    author = {Nishiguchi, Katsuhiko and Ono, Yukinori and Fujiwara, Akira},
    title = {Single-electron counting statistics of shot noise in nanowire Si metal-oxide-semiconductor field-effect transistors},
    journal = {Applied Physics Letters},
    volume = {98},
    number = {19},
    pages = {193502},
    year = {2011},
    month = {05},
    doi = {10.1063/1.3589373},
}

@article{Nishiguchi2014Nanotech,
doi = {10.1088/0957-4484/25/27/275201},
year = {2014},
month = {jun},
publisher = {IOP Publishing},
volume = {25},
number = {27},
pages = {275201},
author = {Katsuhiko Nishiguchi and Yukinori Ono and Akira Fujiwara},
title = {Single-electron thermal noise},
journal = {Nanotechnology},

}

@article{Oono1998,
   author = {Oono, Y. and Paniconi, M.},
   title = {Steady state thermodynamics},
   journal = {Prog. Theor. Phys. Supp.},
   year = {1998},
   volume = {130},
   pages = {29-44},

}

@article{Pekola2012,
   author = {Pekola, J. P. and Saira, O. P.},
   title = {Work, Free Energy and Dissipation in Voltage Driven Single-Electron Transitions},
   journal = {J. Low. Temp. Phys.},
   year = {2012},
   volume = {169},
   pages = {70-76},
   number = {1-2},
   DOI = {10.1007/s10909-012-0659-7},
}

@article{Proesmans2020,
  title = {Finite-Time Landauer Principle},
  author = {Proesmans, Karel and Ehrich, Jannik and Bechhoefer, John},
  journal = {Phys. Rev. Lett.},
  volume = {125},
  issue = {10},
  pages = {100602},
  numpages = {6},
  year = {2020},
  month = {Sep},
  publisher = {American Physical Society},
  doi = {10.1103/PhysRevLett.125.100602},
  url = {https://link.aps.org/doi/10.1103/PhysRevLett.125.100602}
}

@article{Raz2016,
   author = {Raz, O. and Suba\ifmmode \mbox{\c{s}}\else \c{s}\fi{}\ifmmode \imath \else \i \fi{}, Y. and Jarzynski, C. },
   title = {Mimicking Nonequilibrium Steady States with Time-Periodic Driving},
   journal = {Phys. Rev. X},
   year = {2016},
   volume = {6},
   number = {2},
  pages = {021022},
   DOI = {10.1103/PhysRevX.6.021022},
}

@article{Saira2012,
   author = {Saira, O. P. and Yoon, Y. and Tanttu, T. and Möttönen, M. and Averin, D. V. and Pekola, J. P.},
   title = {Test of the {Jarzynski} and {Crooks} Fluctuation Relations in an Electronic System},
   journal = {Phys. Rev. Lett.},
   year = {2012},
   volume = {109},
   pages = {180601},
   number = {18},
   DOI = {10.1103/PhysRevLett.109.180601},
}

@article{Salhani2024,
    author = {Sahlhani, C. and Chida, K. and Shimizu, T. and Hayashi, T. and Nishiguchi. K.},
    title = "{Alternating-current signal sensing beyond cutoff frequency using a single-electron dynamic random access memory}",
    journal = {Phys. Rev. Appl.},
    volume = {23},
    number = {2},
    pages = {L021001},
    year = {2025},
    month = {02},
    doi = {10.1103/PhysRevApplied.23.L021001},
}

@article{Seifert2005,
  title = {Entropy Production along a Stochastic Trajectory and an Integral Fluctuation Theorem},
  author = {Seifert, Udo},
  journal = {Phys. Rev. Lett.},
  volume = {95},
  issue = {4},
  pages = {040602},
  numpages = {4},
  year = {2005},
  month = {Jul},
  publisher = {American Physical Society},
  doi = {10.1103/PhysRevLett.95.040602},
}

@article{Shiraishi2016,
   author = {Shiraishi, N. and Saito, N. and Tasaki, H.},
   title = { Universal Trade-Off Relation between Power and Efficiency for Heat Engines},
   journal = {Phys. Rev. Lett.},
   year = {2016},
   volume = {117},
   pages = {190601},
   number = {19},
   DOI = {10.1103/PhysRevLett.117.190601},
}

@article{Speck2005,
doi = {10.1088/0305-4470/38/34/L03},
year = {2005},
month = {aug},
publisher = {},
volume = {38},
number = {34},
pages = {L581},
author = {T Speck and U Seifert},
title = {Integral fluctuation theorem for the housekeeping heat},
journal = {J. Phys. A: Math. Gen.},
}

@article{Trepagnier2004,
author = {E. H. Trepagnier  and C. Jarzynski  and F. Ritort  and G. E. Crooks  and C. J. Bustamante  and J. Liphardt },
title = {Experimental test of Hatano and Sasa's nonequilibrium steady-state equality},
journal = {Proceedings of the National Academy of Sciences},
volume = {101},
number = {42},
pages = {15038-15041},
year = {2004},
doi = {10.1073/pnas.0406405101},
URL = {https://www.pnas.org/doi/abs/10.1073/pnas.0406405101},
eprint = {https://www.pnas.org/doi/pdf/10.1073/pnas.0406405101},
}

@article{Yoshimura2023,
   author = {Yoshimura, Kohei and Kolchinsky, Artemy and Dechant, Andreas and Ito, Sosuke},
   title = {Housekeeping and excess entropy production for general nonlinear dynamics},
   journal = {Phys. Rev. Research},
   year = {2023},
   volume = {5},
   number = {1},
   pages = {013017},
   DOI = {10.1103/PhysRevResearch.5.013017},
}

@article{Wolpert2019,
doi = {10.1088/1751-8121/ab0850},
url = {https://dx.doi.org/10.1088/1751-8121/ab0850},
year = {2019},
month = {apr},
publisher = {IOP Publishing},
volume = {52},
number = {19},
pages = {193001},
author = {David H Wolpert},
title = {The stochastic thermodynamics of computation},
journal = {J. Phys. A},
}

\end{document}